# Clinical use and future requirements of relative biological effectiveness: survey among all European proton therapy centres

Lena Heuchel[1], Christian Hahn[1,2,3], Jörg Pawelke[2,4], Brita Singers Sørensen[5,6], Manjit Dosanjh [7,8], Armin Lühr[1],*

[1] Department of Physics, TU Dortmund University, Dortmund, Germany

[2] OncoRay – National Center for Radiation Research in Oncology, Faculty of Medicine and University Hospital Carl Gustav Carus, Technische Universität Dresden, Helmholtz-Zentrum Dresden-Rossendorf, Dresden, Germany

[3] Department of Radiotherapy and Radiation Oncology, Faculty of Medicine and University Hospital Carl Gustav Carus, Technische Universität Dresden, Dresden, Germany

[4] Helmholtz-Zentrum Dresden-Rossendorf, Institute of Radiooncology – OncoRay, Dresden, Germany

[5] Department of Experimental Clinical Oncology, Aarhus University Hospital, Aarhus, Denmark

[6] Danish Center for Particle Therapy, DCPT, Aarhus University Hospital, Aarhus, Denmark

[7] Department of Physics, University of Oxford, Oxford, UK

[8] CERN, Geneva, Switzerland

* Corresponding Author Name, Address & Email Address

Armin Lühr,

Otto-Hahn-Straße 4

44227 Dortmund, Germany

Armin.Luehr@tu-dortmund.de

**Abstract**

*Background and purpose:* The relative biological effectiveness (RBE) varies along the treatment field. However, in clinical practice, a constant RBE of 1.1 is assumed, which can result in undesirable side effects. This study provides an accurate overview of current clinical practice for considering proton RBE in Europe.

*Materials and Methods:* A survey was devised and sent to all proton therapy centres in Europe that treat patients. The online questionnaire consisted of 39 questions addressing various aspects of RBE consideration in clinical practice, including treatment planning, patient follow-up and future demands.

*Results:* All 25 proton therapy centres responded. All centres prescribed a constant RBE of 1.1, but also applied measures (except for one eye treatment centre) to counteract variable RBE effects such as avoiding beams stopping inside or in front of an organ at risk and putting restrictions on the minimum number and opening angle of incident beams for certain treatment sites. For the future, most centres (16) asked for more retrospective or prospective outcome studies investigating the potential effect of the effect of a variable RBE. To perform such studies, 18 centres asked for LET and RBE calculation and visualisation tools developed by treatment planning system vendors.

*Conclusion:* All European proton centres are aware of RBE variability but comply with current guidelines of prescribing a constant RBE. However, they actively mitigate uncertainty and risk of side effects resulting from increased RBE by applying measures and restrictions during treatment planning. To change RBE-related clinical guidelines in the future more clinical data on RBE are explicitly demanded.

**Introduction**

Irradiation with protons has two important advantages over irradiation with photons. Firstly, proton therapy offers the possibility of sparing normal tissue much better than conventional radiotherapy, due to its physical dose profile, while maintaining the same dose in the tumour and secondly, proton irradiation is biologically more effective than photon irradiation [1]. Therefore, dose prescription in clinics uses the relative biological effectiveness (RBE) to convert the absorbed dose to an RBE-weighted dose corresponding to an isoeffective photon dose. Thus, it is possible to benefit from many years of clinical experience with photon irradiation, as tolerance doses and prescription doses can still be used [2].

In clinical practice, a constant RBE of 1.1 is currently used. This constant value is based on in vivo studies, is recommended worldwide [3] and assumes proton irradiation to be 10% more effective than photon irradiation. However, the RBE varies along the treatment field, due to the correlation with the linear energy transfer (LET). There is evidence for an increased RBE at the distal edge of a spread-out Bragg peak (SOBP), where many stopping protons result in an elevated LET [3–6]. In addition, RBE depends on other parameters, such as tissue type and dose [7]. Therefore, most models for calculating the RBE of protons are based on the LET, but may also consider various other parameters, e.g. the radiation sensitivity of the tissue [8–11]. The large number of different models [12] complicates a comparison of results between different clinics [13].

Current clinical data suggest a correlation between an increased biologically effective dose, i.e., an elevated RBE and the occurrence of clinically relevant side effects [14–16]. A large amount of these data originates from the analysis of image changes in healthy brain tissue on follow-up magnetic resonance imaging after proton therapy of brain tumour patients [14–19]. But also for other anatomical regions, such as chest-wall patients [20] and breast cancer patients [21], the occurrence of radiographic changes and an increased rate of rib fractures were correlated with elevated LET or RBE levels, respectively. In contrast, there may also be RBE values lower than 1.1 in the tumour [22].

The central question is whether and to which extent the variability of the proton RBE is of clinical relevance. Firstly, in large parts of the irradiated volume, such as the clinical target volume, the variation of the RBE remains small. Secondly, the highest RBE values are often found where the absorbed dose is low and, therefore, no adverse radiation effect is induced. At the same time, the growing awareness of uncertainties in proton therapy caused by possible RBE effects [3,23,24] lets proton therapy centres already apply various methods during treatment planning to mitigate uncertainties in RBE weighted dose, including the avoidance of certain beam angles or consideration of LET distributions in patient plans. The impact of RBE uncertainty on everyday clinical practice was discussed at a workshop by the European Particle Therapy Network (EPTN) work package 6 (WP6) on radiobiology [25]. The discussions emphasized the inevitable insecurity of proton clinics on how to effectively account for RBE variability, which urges a deeper insight into this issue and, particularly, a robust review examining the current clinically applied RBE strategies at a large number of centres.

The current study presents a detailed overview on the actual clinically applied approaches and future needs to consider proton RBE. It is based on a survey on clinical RBE considerations designed by EPTN WP6 and sent to all European proton therapy centres.

**Material and Methods**

*Study design and proton centre selection*

In 2020, a questionnaire on the current clinical practice of RBE considerations was designed by the EPTN WP 6 so it could be distributed among all the proton therapy centres in Europe treating patients. The European Society for Radiotherapy and Oncology (ESTRO) was consulted to provide a list with contact details of all European proton therapy centres. In addition, the list of facilities in clinical operation published by the Particle Therapy Co-Operative Group (PTCOG) [26] was used to identify suitable centres. In total, 25 proton therapy centres from 14 European countries were identified (Fig. 1). Both the clinical heads of the medical and physics departments of each centre were contacted and asked to nominate a contact person who should answer the questionnaire on behalf of the centre. Both a link to the online questionnaire and a copy (cf. Supplement) were sent to the appointed persons.

*Questionnaire design*

The survey was designed as an online questionnaire, which was implemented and answered using SurveyMonkey (SurveyMonkey, San Mateo, USA). The questionnaire consisted of eight sections (cf. Supplement). The first section asked for contact details of the participants and the last offered the possibility for additional comments. Accordingly, the main part of the survey addressing the issue of RBE consideration in the clinic consisted of 24 questions with predefined answers (closed questions) and 15 questions with free text answers (open questions), divided into six different sections within the following topic headings (Table 1):

- o Treatment field arrangement
- o Robust optimization

- Variability of RBE in treatment planning
- Estimation of LET and RBE for patient treatment
- RBE consideration for patient follow-up
- Future improvements

Some questions allowed the selection of more than one answer option and if desired, individual questions could be skipped. Therefore, some questions have not been answered by all participants. In addition, participants had the opportunity to add comments to most of the questions.

*Analysis of responses*

Different evaluation and analysis strategies were used for closed and open questions. The answers to the closed questions were evaluated quantitatively. A more qualitative study analysis was performed for evaluating the open questions looking for similarities and considering the context in the given answers. Answers that were given in a similar way by different centres were identified. Thereby, responses could be categorized into groups and then also evaluated quantitatively. In some cases, answers to topic-related questions were considered together and their results were summarised.

## Results

All 25 European proton therapy centres responded to the questionnaire between June 2020 and May 2021. In 15 centres, the questions were answered by physicists, in nine by physicians, and in one by both together. Not all centres answered all 39 questions. The response rate differed depending on the type of question. On average, individual open and closed questions were answered by 54% and 98% of participating centres, respectively (Table 1). One participating centre was treating only eye tumours making several questions not appropriate for them. For this reason, it was excluded from the following analysis and 24 centres were considered as 100%, if not stated otherwise.

Almost all participating centres treated base of skull (92%), brain (92%) and head and neck tumours (88%) (cf. Supplement Fig. S1). Moreover, prostate, lung and craniospinal irradiations were performed by 42%, 50% and 71% of the centres, respectively. All other examined treatment sites (oesophagus, breast, liver, pancreas) were treated by less than 38% of the centres.

All centres reported to follow current guidelines on proton dose prescription and prescribed a constant RBE of 1.1 (Fig. 2a). However, at the same time, all centres applied measures to actively counteract uncertainties resulting from a variable RBE (Fig. 2b). Most centres (71%) actively considered the possibility of a variable RBE only for organs at risk (OAR) but none for the target volume alone (Fig. 3a). To counteract a variable RBE, nearly all centres used special beam arrangements (96%) or avoided beams stopping in an OAR (100%, Fig. 3b).

In the course of treatment planning, all centres considered at least some restrictions on the arrangement of treatment fields (Table 2). The most frequent restriction concerned the choice of incident beam direction. All centres tried to exclude beam angles, where a beam stops in front of or inside an OAR. Where this was unfeasible, nine centres applied, in addition, lower field weights. Of these nine, four centres additionally extended the field range to position the end of the proton track beyond an OAR ("shoot-through").

The minimum number of beam orientations depended at most centres (92%) on the treatment site and differed between centres. Two or more beam orientations were used for almost all treatment sites. Treatment plans with only one beam were never used for certain tumour sites (base of skull, oesophagus, pancreas, prostate) while they were rarely allowed (less than 15% of centres) for other sites (Table S1). The angle between two proton beams constituted another commonly used restriction. Three-quarters of the centres tried to avoid having two beams originating from similar directions and thus stopping in the same anatomical area. Ten centres precisely specified their angle restriction. The beams were required to be separated by at least 10°, at least 30° or even larger angles by two, six and two centres, respectively.

Nearly all centres (88%) used robust optimization for treatment planning. Most centres (83%) considered robustness of target coverage in the optimisation process. Half of the centres considered, in addition, robustness of dose limits in selected OARs, while some (17%) considered robust dose objectives for all OAR. Three centres never applied robust optimisation.

The vast majority of centres (88%) performed at least some patient-specific calculations of, e.g., LET or RBE. However, the results of these calculations are rarely used to support patient treatment, but mostly for research purpose. Only twelve centres had already used these calculations at least at some point in time to support treatment planning. For clinical treatment, most commonly, the LET distribution was calculated (42%) followed by the variable RBE distribution (25%, Fig. 4a). Such patient-specific calculations of different RBE-related quantities were most often (11 centres) used for retrospective analyses. Ten centres used them for plan evaluation and ten for clinical research. Only three centres performed these calculations during treatment planning. Thus, LET or RBE was mostly calculated for research purposes.

The frequency of performing calculations of, e.g., LET and RBE differed between centres and also depended on the tumour entity. Most frequently, calculations were performed for brain (33%) and head and neck tumour patients (17%). To perform these calculations, the largest group of centres (38%) used a research version of the treatment planning system (TPS) RayStation (RaySearch, Stockholm, Sweden), while others applied different research Monte-Carlo simulation or in-house software frameworks (Fig. 4b). A calculation was most often initiated by physicists, followed by the treating physician and mostly performed by physicists or research staff.

More than half of the centres (58%) considered RBE during patient follow-up. Most centres (42%) intended to better understand the patient's radiation response. Only eight centres specified how RBE was considered during follow-up: If side effects occurred, then investigations were performed to determine whether these can be explained by the RBE distribution. RBE was mainly considered for brain and head and neck tumour patients, as specified by nine centres.

The survey's part on future improvements had a response rate of 68%, which was higher than the average for open questions of the other topics (40%, Table 1). In general, the most frequent and urgent request of the centres was for more clinical evidence for possible effects due to an increased RBE. Most centres (67%) would like to see more published retrospective and prospective studies with large patient cohorts examining the impact of RBE on the probability of toxicity occurrence. To generate such clinical evidence, 18 centres indicated that they would like to be able to visualize LET and RBE for all patients. Nine centres would like to apply LET and RBE visualization beyond research purposes to support treatment planning, e.g., during plan approval. Therefore, the most sought-after tool that should be developed by vendors is a visualization of LET and RBE in their clinical TPS. In addition, participants from seven centres – of which six are physicists and one is a physician – indicated that they would like vendors to develop LET and RBE optimization tools as well.

There was no consensus on the issue of possible (future) guidelines in the context of a variable proton RBE. Five participants expressed that at this point in time there cannot be any guidelines due to the lack of clinical evidence. Four other centres proposed guidelines for RBE-related (physical) quantities. These four centres asked for harmonization of the calculation of variables, such as LET, and they suggested to adopt these calculations as standard practice for all patients, mainly for research purposes, but also to support treatment planning.

**Discussion**

Initial clinical evidence indicates an increased RBE at the distal end of the SOBP, which may lead to undesirable and unexpected side effects after proton therapy [14–17,19–21]. A survey to study to which extent the RBE variability is already considered in daily clinical routine was sent to 25 proton therapy centres in Europe. The high response rate of 100% underlines the relevance of RBE variability. On the one hand, all centres followed current guidelines [2] and prescribed a constant RBE of 1.1. On the other hand, all centres applied measures to counteract the variability of the RBE, such as avoidance of angles where the beam stops inside or in front of an OAR, avoidance of acute angles between incident beams, use of more fields to reduce the relative weight of critical incident angles or extending the field range to position the end of the proton track beyond an OAR, except for the centre performing eye irradiation only. These measures are adopted since many years in proton therapy planning as they not only address the issue of a potentially variable RBE but also account for range uncertainties. Nevertheless, the answers to our survey demonstrate that all proton therapy centres are well aware of the issue of a variable RBE and are actively trying to minimize possible (adverse) effects of the variable RBE.

The current strategy of maintaining target dose prescription using a constant RBE while mitigating the impact of variable RBE in OAR is consistent with a recent AAPM report [3] and research works on optimising LET/RBE only in OAR [19,28–30]. The RBE effect is expected to be highest at the distal edge of the SOBP, which is usually placed in healthy tissue due to mandatory range uncertainty margins [5]. Accordingly, follow-up studies found radiation-induced image changes at the distal field edges, where high LET spatially coincides with high dose in radiosensitive OAR [14–21].

The reported clinical strategies were shown to be effective in reducing variable RBE weighted dose hotspots in critical OAR in some cases in recent research works [31,32].

However, a replanning study of glioma patients [31] showed that solely increasing the number of treatment fields may in some cases be insufficient to reduce the risk of side-effects resulting from an increased biologically effective dose. Furthermore, it was found that extending the proton range beyond an OAR ("shoot-through") results in an increased volume receiving a high biological dose without reducing the maximum radiobiological effective dose to the OAR [32]. Whether and how effective the clinically applied measures are, thus, depends on the individual patient case. The fact that centres are already using countermeasures could influence clinical outcomes, as fewer RBE related side effects have been reported than would have been expected based on estimated RBE values from preclinical studies. An exact assessment of the benefit of the applied clinical measures requires LET/RBE calculations tools for treatment plans in the clinical environment. Accordingly, almost all centres request LET and RBE visualisation tools provided by the TPS vendors.

The question of how to proceed regarding RBE is of great concern to the proton therapy centres in Europe. They emphasized the need for clinical data on RBE variability with clinical relevance beyond those few retrospective studies known to them [14–17]. Especially, changing the current guidelines of prescribing a constant RBE of 1.1 requires many more retrospective or ideally prospective studies that correlate clinical outcome with RBE or other RBE-related quantities, such as LET. This is in line with recently published recommendations for consistent toxicity scoring and follow-up in adult brain tumour patients, which state that radiological follow-up for image changes after radiotherapy are "considered minimum of care" [33]. Many centres are also requesting guidelines to harmonize calculation and reporting of LET and RBE values to ensure comparability of studies and treatment outcomes. Recently, it was shown that a harmonization of LET and RBE calculation among centres is possible and would also allow for consistent multi-centric analysis of toxicity caused by an increased RBE [34].

The tumour entities treated by most participating centres are brain and head and neck tumours, which is in line with a study on the current practice in proton therapy delivery in Europe [35]. In brain tumour patients, radiosensitive OAR are usually situated close to the target volume, resulting in areas being exposed to high dose and high LET and, therefore, at risk of more pronounced clinical effects due to an elevated RBE dose. Furthermore, most LET and RBE calculations have been performed for these anatomical sites. Hence, for these anatomical regions, most evidence for clinically relevant variable RBE effects is available [14–17] and more may be expected in the future. Additionally, pronounced local RBE effects may result from increasing precision in proton therapy through the reduction of range and setup uncertainty [36], as a smaller smear-out effect of the RBE is expected [6].

For the centre treating only eye tumours, several questions were inadequate due to their specialized treatments. This centre stated that possible RBE effects are less important for ocular irradiation as no high-risk organs are located near the target volume. Nevertheless, an ocular TPS taking a variable RBE into account was considered to be of a great value.

This dedicated survey provided detailed insights into the current clinical practice regarding the consideration of RBE at proton therapy centres in Europe. Possible effects of a variable RBE are already being countered by all European centres through initial measures, especially through the arrangement of treatment fields. In this way, the uncertainty and risk of side effects resulting from increased RBE can be reduced, while complying with the current guidelines of a constant RBE. However, centres insistently call for more clinical data, e.g., retrospective studies correlating clinical outcome with the biologically effective dose, in particular, to serve as a basis for changing RBE-related clinical guidelines. Enabling proton therapy centres to calculate LET and RBE in their clinical setting is considered as the next crucial step to overcome the RBE issue.

**Acknowledgements**

The authors are grateful to Prof. Aswin Hoffman for his advice in designing the survey and Prof. Peter van Luijk, Dr. Anne Vestergaard and Dr. Dirk Wagenaar for critically proofreading the questionnaire. In addition, ESTRO's assistance in identifying suitable centres and contact addresses is acknowledged. The authors would like to express their special thanks to all participating centres and the colleagues who completed the questionnaires.

**Tables:**

**Table 1**: Design of the questionnaire: for each topic, the number of closed, open and all questions is provided. The stated relative response rates assume a total of 24 participating centres.

**Table 2**: Number of centres applying certain restrictions or measures to counteract a potentially variable RBE.

**Figures**:

**Fig. 1:** The locations of the 25 proton therapy centres from 14 European countries that participated in the survey are marked with red dots on a map [27].

This map is made available under the Open Database License and any rights in individual contents of the database are licensed under the Database Contents License (both: http://opendatacommons.org/licenses/dbcl/1.0/)

**Fig. 2:** (a) Responses to the question: Do you prescribe anything else than a fixed relative biological effectiveness (RBE) of 1.1 for patient treatment? (b) Number of centres applying measures to counteract a potentially variable RBE.

**Fig. 3:** Responses to the questions: (a) Where do you actively consider the possibility of a variable relative biological effectiveness (RBE)? (b) Which kind of measures do you apply to consider/counteract a potentially variable proton RBE? (LET: linear energy transfer)

**Fig. 4:** Responses to questions concerning the topic estimation of LET and RBE for patient treatment: (a) Which quantities do you calculate for clinical treatments? (b) Which software do you use to perform these calculations? (open question) (LET: linear energy transfer, RBE: relative biological effectiveness, NTCP: normal tissue complication probability)

| Topic | Closed questions | | Open questions | | Total | |
|---|---|---|---|---|---|---|
| | Number | Response rate | Number | Response rate | Number | Response rate |
| Treatment field arrangement | 7 | 98.3% | 2 | 72.0% | 9 | 92.4% |
| Robust optimization | 2 | 96.0% | 0 | - | 2 | 96.0% |
| Variability of RBE in Treatment planning | 6 | 96.7% | 1 | 12.0% | 7 | 84.6% |
| Estimation of LET and RBE for patient treatment | 6 | 99.3% | 5 | 42.4% | 11 | 73.5% |
| RBE consideration for patient follow-up | 3 | 100.0% | 1 | 33.3% | 4 | 50.0% |
| Future improvements | 0 | - | 6 | 68.0% | 6 | 68.0% |
| Total | 24 | 98.0% | 15 | 54.0% | 39 | 79.5% |

| Restriction / measure | Number of centres | Percentage of centres |
|---|---|---|
| Minimum number of beam orientations | 22 | 92% |
| Avoiding beams that stop inside / in front of OAR | 24 | 100% |
| Minimum angle between two beams | 18 | 75% |
| Lower field weights for beams stopping inside / in front of OAR | 9 | 38% |
| "Shoot-through" | 4 | 17% |

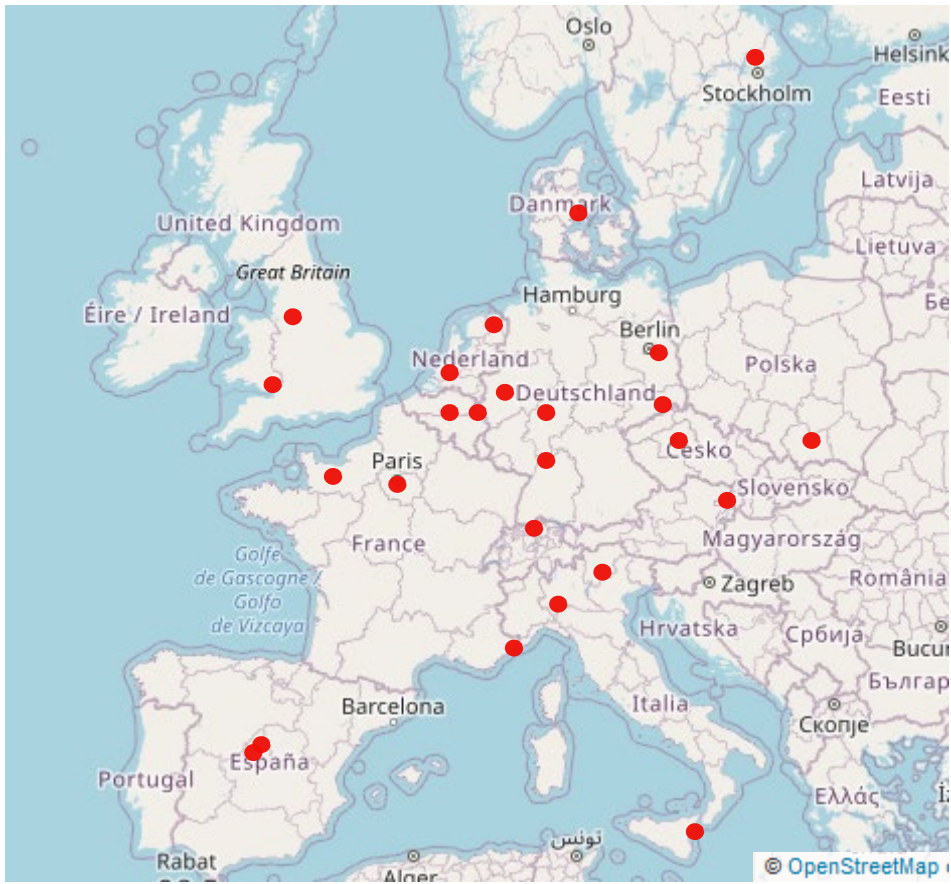

**Fig. 1:** The locations of the 25 proton therapy centres from 14 European countries that participated in the survey are marked with red dots on a map [27].

This map is made available under the Open Database License and any rights in individual contents of the database are licensed under the Database Contents License (both: http://opendatacommons.org/licenses/dbcl/1.0/)

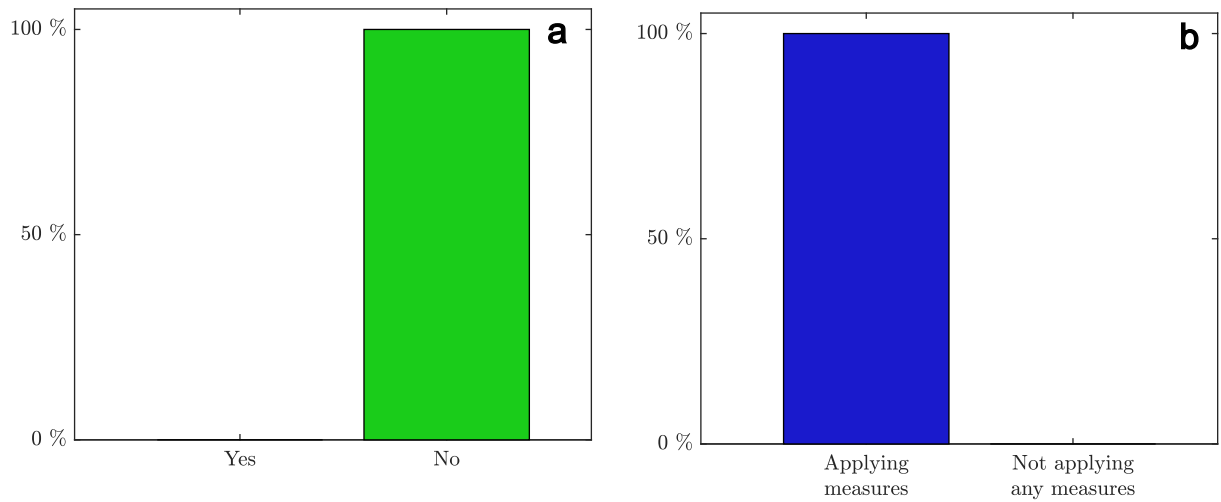

**Fig. 2:** (a) Responses to the question: Do you prescribe anything else than a fixed relative biological effectiveness (RBE) of 1.1 for patient treatment? (b) Number of centres applying measures to counteract a potentially variable RBE.

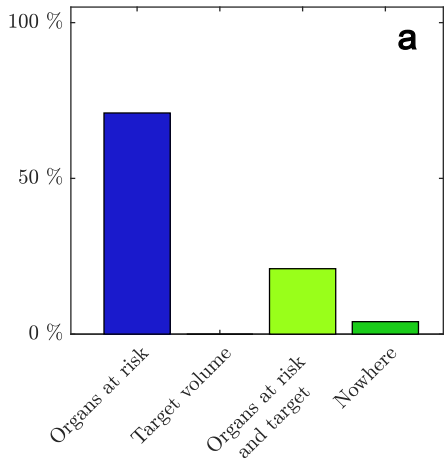
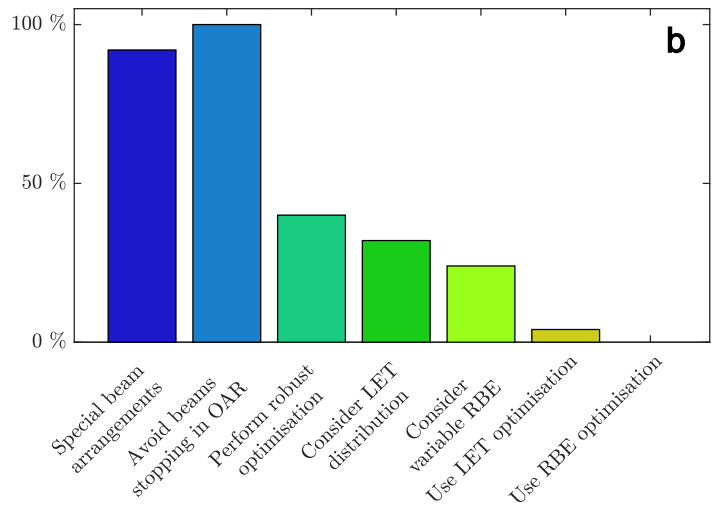

**Fig. 3:** Responses to the questions: (a) Where do you actively consider the possibility of a variable relative biological effectiveness (RBE)? (b) Which kind of measures do you apply to consider/counteract a potentially variable proton RBE? (LET: linear energy transfer)

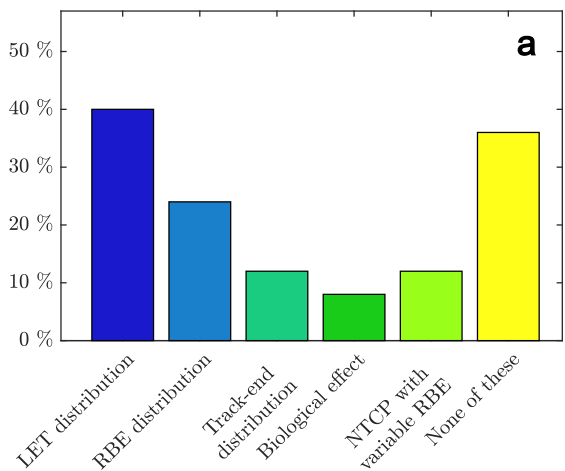
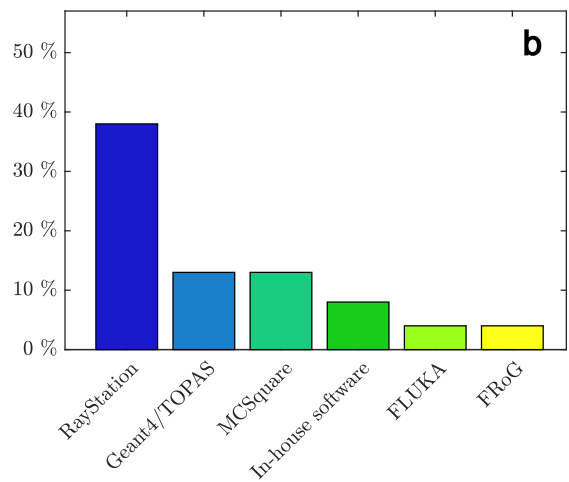

**Fig. 4:** Responses to questions concerning the topic estimation of LET and RBE for patient treatment: (a) Which quantities do you calculate for clinical treatments? (b) Which software do you use to perform these calculations? (open question) (LET: linear energy transfer, RBE: relative biological effectiveness, NTCP: normal tissue complication probability)